\documentclass[10pt]{article}
\usepackage{amsmath, amssymb}
\usepackage[OE]{expresss}

\usepackage{graphicx}% Include figure files

\newcommand\ee{\mathrm{e}}

\newcommand\dd{\, \mathrm{d}}
\newcommand\ii{\mathrm{i}}
\newcommand\lt{\left}
\newcommand\rt{\right}

\newcommand\nm{\textrm{~nm}}
\newcommand\um{\mathrm{~\mu m}}

\begin{document}

\title{First-order optical spatial differentiator\\
 based on a guided-mode resonant grating}

\author{Dmitry~A.~Bykov,\!\authormark{1,2,*} Leonid~L.~Doskolovich,\!\authormark{1,2}\\ Andrey~A.~Morozov,\!\authormark{1,2} Vladimir~V.~Podlipnov,\!\authormark{1,2} \\ Evgeni~A.~Bezus,\!\authormark{1,2} Payal~Verma,\!\authormark{2,3} and Victor~A.~Soifer\authormark{1,2}}
\address{\authormark{1}Image Processing Systems Institute --- Branch of the Federal Scientific Research Centre ``Crystallography and Photonics'' of Russian Academy of Sciences, 151 Molodogvardeyskaya st., Samara 443001, Russia\\
\authormark{2}Samara National Research University, 34 Moskovskoye shosse, Samara 443086, Russia\\
\authormark{3}Department of Electronics and Communication Engineering, Dayananda Sagar University, Bangalore 560068, India}
\email{\authormark{*}bykovd@gmail.com}

\begin{abstract}
We present an experimental demonstration of a subwavelength diffraction grating performing first-order differentiation of the transverse profile of an incident optical beam with respect to a spatial variable.
The experimental results are in a good agreement with the presented analytical model suggesting that the differentiation is performed in transmission at oblique incidence and is associated with the guided-mode resonance of the grating.
According to this model, the transfer function of the grating in the vicinity of the resonance is close to the transfer function of an exact differentiator.
We confirm this by estimating the transfer function of the fabricated structure on the basis of the measured profiles of the incident and transmitted beams.
The considered structure may find application in the design of new photonic devices for beam shaping, optical information processing, and analog optical computing.
\end{abstract}


%\bibliographystyle{osajnl}
%\bibliography{Experiment}

\section{Introduction}

Optical devices performing specific temporal and spatial transformations of optical signals are of great interest for a wide range of applications including all-optical information processing and analog optical computing. 
For optical information processing in the spatial domain, bulky optical systems containing lenses and optical filters are commonly utilized~\cite{Goodman:1968}.
One of the most widely used systems of this type is the optical correlator containing a spatial filter and one or two lenses performing Fourier transforms. 
The complex transmission coefficient of the filter defines the transfer function of the system and describes the transformation of the incident optical signal in the Fourier domain.

Recent developments in the field of nanophotonics and metamaterials enable creating ultracompact optical devices working as an optical correlator and having the thickness comparable to the wavelength of the incident light. 
Generally speaking, the approaches to the design of nanophotonic structures for optical differentiation or integration can be divided into two groups. In the first group (metasurface approach), an optical correlator with graded-index lenses is used, in which the spatial filter comprises a metasurface (an array of nanoresonators with varying parameters) encoding the complex transmission coefficient of a differentiating or integrating filter~\cite{Silva:2014:science, Pors:2015:nl}. 
In particular, in a recent work~\cite{Silva:2014:science} it was theoretically shown that a metamaterial layer with a specially designed structure enables performing the operations of differentiation and integration of an optical beam with respect to a spatial coordinate. 
In the second group (resonant structure approach), various resonant diffractive structures including systems of multiple homogeneous layers~\cite{my:Bykov:2014:ol, my:Bykov:2014:oe, Zangeneh:2018:oc, Zhu:2017:nc, Youssefi:2016:ol, Zangeneh:2017:ol, Ruan:2015:ol}, plasmonic circuits~\cite{Hwang:2016:apl, Hwang:2018:oe}, and resonant diffraction gratings~\cite{my:Golovastikov:2014:qe:eng, Guo:2018:opt, Saba:2017:arxiv} are used. 
The possibility to utilize such resonant structures for optical differentiation or integration is based on the fact that the Fano profile describing the reflection or the transmission coefficient of the structure in the vicinity of the resonance can approximate the transfer function of a differentiating or integrating filter~\cite{my:Bykov:2014:ol, my:Bykov:2014:oe, my:Golovastikov:2014:qe:eng}.

The present authors believe that the resonant structure approach has several advantages over the metasurface approach. In particular, optical differentiators and integrators based on resonant structures are much more compact since they do not require the utilization of additional lenses implementing the Fourier transform. In addition, these structures are much easier to fabricate than metasurfaces, which, as mentioned above, usually consist of a large number of nanoresonators with different parameters.

Despite a large number of theoretical papers, the results of experimental studies of spatial differentiators and integrators are presented only in a few works. First experimental results on the fabrication and investigation of a metasurface-based differentiator and integrator consisting of arrayed gold nanobricks were presented in 2015 in~\cite{Pors:2015:nl}.

Differentiating and integrating metasurfaces of~\cite{Silva:2014:science, Pors:2015:nl} are used in conjunction with a lens. Also, due to high technological complexity of metasurface fabrication, the transfer function of the differentiating filter was implemented with a significant error, see Fig.~4(c) in~\cite{Pors:2015:nl}. As a result, the implementation of differentiation was not demonstrated experimentally. Nevertheless, the structure made it possible to implement edge detection, albeit with a relatively low quality, see Fig.~6(e) in~\cite{Pors:2015:nl}.

In~\cite{Zhu:2017:nc}, a differentiator based on the Kretschmann configuration was proposed. In this case, the spatial differentiation is performed due to a resonant effect associated with excitation of a surface plasmon at a single metal--dielectric interface. 
This structure can be easily fabricated, since it consists of homogeneous layers and does not require lithography.
Experimental results presented in~\cite{Zhu:2017:nc} demonstrate high-quality edge detection of an image. 
At the same time, the accuracy of the derivative calculation was not investigated in~\cite{Zhu:2017:nc}.
Moreover, the differentiator presented in~\cite{Zhu:2017:nc} contains a bulky prism and therefore has relatively large dimensions.

In the present work, we for the first time present experimental studies of a differentiator based on a guided-mode resonant diffraction grating. 
We show that the presented experimental results are in good agreement with the proposed theoretical description, demonstrating the ability of resonant diffraction gratings to perform first-order differentiation of optical beams profiles with respect to one of the spatial variables.

\section{Theoretical description}

\begin{figure}[tbh]
	\centering
	\includegraphics{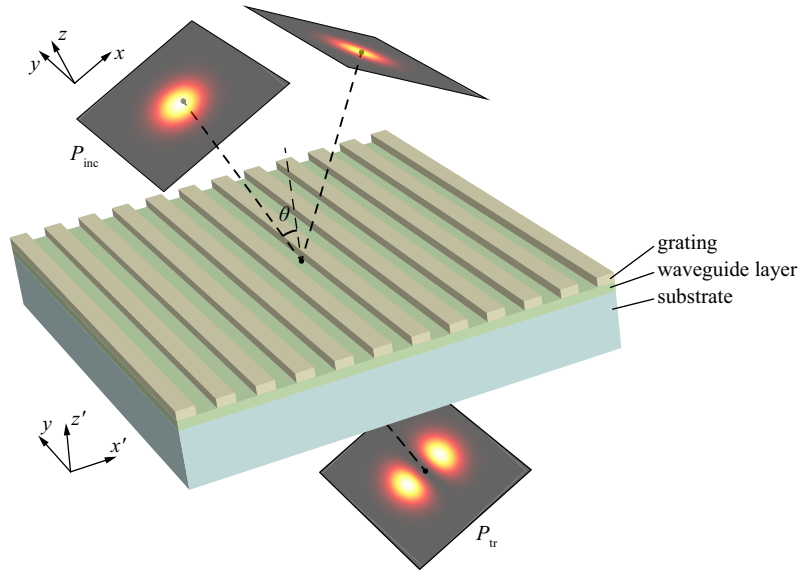}	
	\caption{\label{fig:1}Diffraction of an optical beam on a resonant diffraction structure consisting of a grating on top of a slab waveguide layer deposited on a substrate.
	The transverse profiles of the incident ($P_{\rm inc}$), transmitted ($P_{\rm tr}$), and reflected beams schematically illustrating the spatial differentiation performed in transmission are shown.
	}
\end{figure}

Let us consider a monochromatic optical beam impinging on a subwavelength diffraction grating (Fig.~\ref{fig:1}). We assume that the incident beam is linearly polarized in the plane $x y$, and its profile is described by the function $P_{\rm inc} \left(x, y \right)$, where $x$ and $y$ are the coordinates associated with the incident beam (see Fig.~\ref{fig:1}). Upon diffraction on the grating, reflected and transmitted beams will be formed in zeroth diffraction orders. The profile of the transmitted beam is denoted by $P_{\rm tr} \left(x, y \right)$ in Fig.~\ref{fig:1}.

The transformation of the incident beam profile by a subwavelength diffractive structure can be described in terms of linear system theory~\cite{my:Bykov:2014:oe}.
Assuming that the incident beam is polarized along the $x$ axis (see Fig.~\ref{fig:1}), we can characterize this transformation by the following transfer function (see the Appendix):
\begin{equation}
\label{eq:1}
 H\left(k_x, k_y \right)\approx T\left(k_x \cos \theta + \sqrt{k_0^2 -k_x^2 -k_y^2 } \cdot \sin \theta, k_y \right),
\end{equation}
where $k_0 = \omega/{\rm c}$ is the free-space wave number, $\omega$ is the angular frequency of the incident beam,
$\theta$ is the angle of incidence of the beam (see Fig.~\ref{fig:1}), and $T\left(k_{x'}, k_y \right)$ is the transmission coefficient of the diffraction grating, where $k_{x'}$ is the wave vector component in the grating coordinate system [see Eq.~\eqref{eq:A0} in the Appendix]. 
The transfer function of Eq.~\eqref{eq:1} describes the complex amplitude of the zeroth transmitted diffraction order corresponding to the incident plane wave with the in-plane wave vector components $k_{x'}, k_y$.

In~\cite{my:Bykov:2017:oe}, on the basis of the spatiotemporal coupled-mode theory, it was shown that the following approximate expression for the transmission coefficient of resonant diffractive structures (guided-mode resonant gratings or photonic crystal slabs) is valid:
\begin{equation}
\label{eq:2}
T\left(k_{x'}, k_y \right) = t
\frac{v_{\rm g}^2 k_{x'}^2 -\left(\omega -\omega _{\rm zt} -\eta_{\rm zt} k_y^2 \right)\left(\omega - \omega_{\rm p2} -\eta_{\rm p2} k_y^2 \right)}
{v_{\rm g}^{2} k_{x'}^2 -\left(\omega -\omega_{\rm p1} -\eta_{\rm p1} k_y^2 \right)\left(\omega -\omega_{\rm p2} - \eta_{\rm p2} k_y^2 \right)}.
\end{equation} 
This resonant approximation is valid in the vicinity of the resonance frequency $\omega_{\rm p1}$ for small $k_x$ and $k_y$ (at small angles of incidence).
In the present work, we are primarily interested in the general form of the transmission coefficient and not in the meaning or in the particular values of the parameters in Eq.~\eqref{eq:2}, which are described in detail in~\cite{my:Bykov:2017:oe}. 
However, let us note that in the case of a lossless subwavelength diffraction grating, the parameters $v_{\rm g}, \omega_{\rm zt},$ and $\omega_{\rm p2}$ are real numbers~\cite{my:Bykov:2015:pra, my:Bykov:2017:oe}.
If necessary, the values of all the parameters used in Eq.~\eqref{eq:2} can be rigorously calculated by means of the Fourier modal method and the S-matrix formalism~\cite{my:Bykov:2017:oe}.

Equations~\eqref{eq:1} and~\eqref{eq:2} define the general class of spatiotemporal transformations of optical signals that can be implemented by resonant diffraction gratings. 
In what follows, we show that one of the possible transformations is the differentiation with respect to one of the spatial variables.

Let us consider the transmission coefficient of Eq.~\eqref{eq:2} in the vicinity of the point $k_{y0} = 0$, $k_{x'0} = v_{\rm g}^{-1} \left[\left(\omega - \omega_{\rm zt} \right)\left(\omega - \omega_{\rm p2}\right) \right]^{1/2}$ at fixed frequency $\omega$ of the incident light. 
At this point, $T\left(k_{x'0}, k_{y0} \right) = 0$ and the Taylor expansion of the transmission coefficient with respect to $k_{x'}$ and $k_y$ up to the lowest-order terms reads as
\begin{equation}
\label{eq:3}
T\left(k_{x'}, k_y \right) \approx \alpha \left(k_{x'} - k_{x'0} \right) + \beta k_y^2.
\end{equation} 
Explicit expressions for the quantities $\alpha$ and $\beta$ can be obtained from Eq.~\eqref{eq:2}. 
Let us assume that the incident beam is symmetric and weakly focused, i.e. it has same narrow spectral widths with respect to $k_x$ and $k_y$.
In this case, the second term in Eq.~\eqref{eq:3} can be neglected, and the transmission coefficient becomes
\begin{equation}
\label{eq:4}
T\left(k_{x'}, k_y \right) \approx \alpha \left(k_{x'} -k_{x'0} \right).
\end{equation} 
Under the same assumptions Eq.~\eqref{eq:1} takes the form $H\left(k_x, k_y \right) \approx T\left(k_x \cos \theta + k_0 \sin \theta, k_y \right)$.
According to Eq.~\eqref{eq:4}, if we choose the angle of incidence $\theta$ from the condition $\sin \theta = k_{x'0} / k_0$, the transfer function of the diffraction grating is proportional to the transfer function of the spatial differentiator:
\begin{equation}
\label{eq:5} H\left(k_x, k_y \right)\sim {\rm i}k_x .
\end{equation} 
The tilde notation here means equality up to a multiplicative constant.
This constant, $\alpha' = -\ii \alpha \cos\theta$, defines the energy efficiency of the differentiator (the amplitude of the output signal). 
Let us note that since the transfer function $H$ is dimensionless and $k_x$ has the dimensions ${\rm \mu m}^{-1}$, the constant $\alpha'$ is a dimensional quantity measured in $\rm \mu m$.

Thus, we have shown that a guided-mode resonant diffraction grating with one-dimensional periodicity enables spatial differentiation of an obliquely incident optical beam. 
The differentiation is performed near the resonance frequency in the case of small angles of incidence.

\section{Experimental results}

\subsection{Design and fabrication of the structure}

In order to demonstrate spatial differentiation of optical beams, we fabricated and experimentally investigated a guided-mode resonant grating. 
The grating was patterned into a layer of electron resist ERP-40 deposited on a $\rm{TiO}_2$ slab waveguide on a quartz substrate (Fig.~\ref{fig:1}).
The dielectric permittivity values of the used materials were measured using the ellipsometric approach. 
At the chosen design wavelength $630\nm$, we obtained the following refractive index values: 
 $1.480$ for the grating,
 $2.395$ for $\rm{TiO}_2$ waveguide layer,
 and $1.459$ for the substrate. 

To obtain the parameters of the structure, we optimized them from the condition of maximizing the angular FWHM of the resonant transmission minimum, which provides a better differentiation quality~\cite{my:Bykov:2014:ol}.
At the same time, the transmission coefficient of the grating should vanish at the design wavelength $630\nm$ and the chosen angle of incidence~$\theta = 1^\circ$, providing the differentiation condition of Eq.~\eqref{eq:4}.
The FWHM calculation (the calculation of the grating transmission spectrum) was based on the Fourier modal method~\cite{Moharam:1995:josaa, Li:1996:josaa2}.

In the optimization, the parameters of the structure were subject to the following constraints.
The thickness of the resist layer (grating height) was fixed at $300\nm$, being limited by the used resist deposition technique.
The grating ridge aspect ratio was limited by 2 according to the used lithography process.
The period $d$ of the grating should provide excitation of a guided mode in the $\rm{TiO}_2$ slab by the first diffraction order, resulting in $d > 265\nm$. At the same time, the grating should be subwavelength, i.e. only the zeroth propagating diffraction order should exist in the substrate and superstrate regions, yielding $d < 430\nm$. 
The optimization resulted in the following parameters of the structure:
period $d=410\nm$; waveguide thickness $h_{\rm wg} = 270\nm$; grating height $h_{\rm gr} = 300\nm$; grating groove width $w = 150\nm$.

\begin{figure}[tbh]
	\centering
	\includegraphics{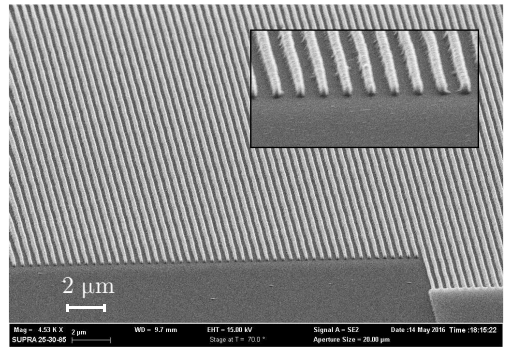}
	\caption{\label{fig:2}Scanning electron microscopy image of the diffraction grating fabricated from ERP-40 electron resist on top of a $\rm{TiO}_2$ layer.}
\end{figure}

The designed guided-mode resonant grating was fabricated as follows.
At the first step, a $\rm{TiO}_2$ slab waveguide was deposited on a quartz substrate using reactive magnetron sputtering with Techport Sputter Coater. 
Then, a diffractive grating was fabricated on the waveguide layer from electron resist ERP-40 using Carl Zeiss Supra 25 electron microscope with e-beam lithography device XeDraw2. 
Typical scanning electron microscopy image of the fabricated grating is shown in Fig.~\ref{fig:2} (for image acquisition, a thin gold layer was deposited on top of the grating).

The parameters of the fabricated grating slightly differ from the design values given above.
However, as it follows from the grating theory, small changes in the parameters result in a shift of the resonance, and not in its disappearance.
Moreover, the condition~\eqref{eq:4} holds provided that the structure is still symmetric and lossless~\cite{my:Bykov:2015:pra}.
Therefore, the imperfections can be compensated by tuning the light frequency and/or the incidence angle $\theta$ of the impinging beam.

We performed numerical simulations to investigate the tolerance of the designed structure to the variation in the geometrical parameters.
For example, $\pm 10\nm$-variation in the parameters result in the shift of the resonance wavelength by no more than $15\nm$ at a fixed angle of incidence.
Let us note that it is the imperfections in the grating period $d$ that affect the resonance wavelength the most.
However, the used lithography technique provides high accuracy in the period of the fabricated structure.
In this case, $\pm 10\nm$-variation in the remaining parameters leads to a resonance wavelength shift not exceeding $5\nm$.

\subsection{Measurement technique}

\begin{figure}[tbh]
	\centering
	\includegraphics{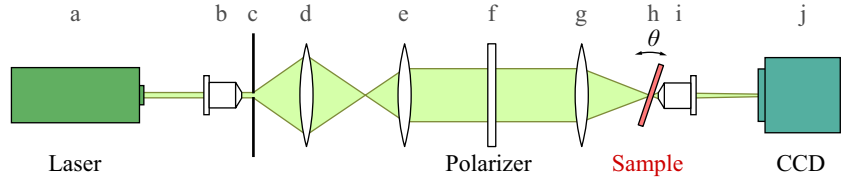}
	\caption{\label{fig:3}Optical setup of the experiment:
(a)~light source, 
(b)~$20\times$ microobjective,
(c)~$40\um$ pinhole,
(d) and~(e)~lenses,
(f)~polarizer, 
(g)~lens with a focal length of 120~mm,
(h)~sample grating,
(i)~$8\times$  microobjective,
(j)~CCD matrix.}
\end{figure}

The optical properties of the fabricated structure were investigated using the following optical setup shown in Fig.~\ref{fig:3}. 
As a light source~(a), tunable laser EKSPLA~NT242 was used.
This laser generates an elliptically-shaped optical beam, which requires filtering.
Using a $20\times$ microobjective~(b), we focused the beam on a pinhole~(c) with $40\um$ aperture, which acted as a secondary point source.
The required beam size was achieved using a collimator consisting of two lenses,~(d) and~(e). After the collimator, the beam passed through a polarizer~(f) used for controlling the polarization of the incident beam with respect to the periodicity direction of the studied grating. Then, the beam was focused onto the sample~(h) by a lens~(g) with a focal length of 120~mm. The sample was mounted so that it could be rotated about a certain axis, being previously centered with respect to this axis using micrometer screws. The sample was imaged on a CCD matrix~(j) with an $8\times$ microobjective~(i).

\subsection{Results and discussion}

\begin{figure}[tbh]
	\centering
	\includegraphics{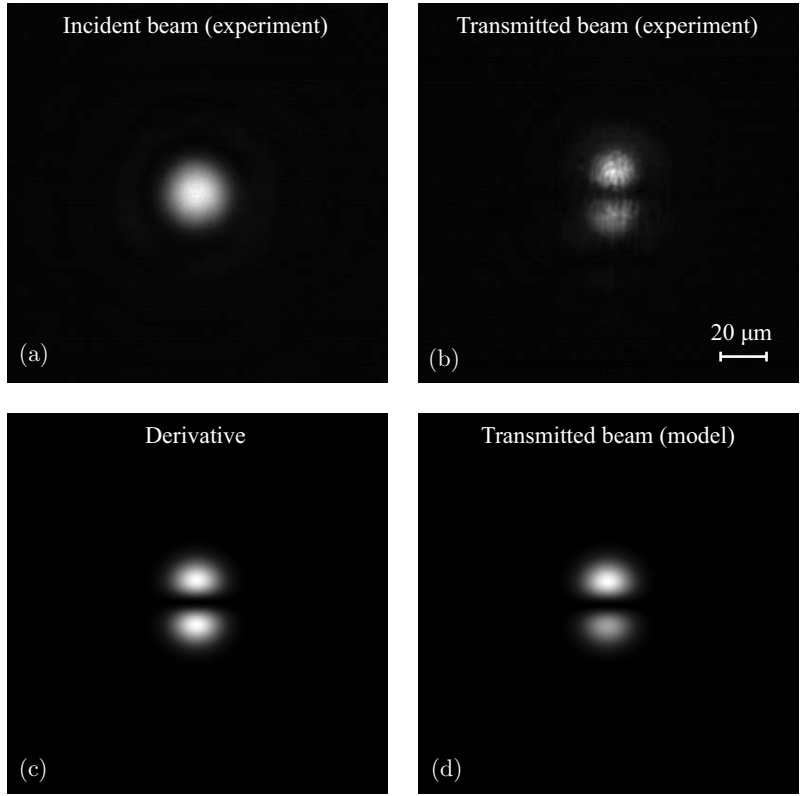}
	\caption{\label{fig:4}(a)~Measured profile of the incident Gaussian beam ($\sigma = 9.52\um$), (b)~measured profile of the transmitted beam, (c)~analytically calculated derivative of the  incident beam, (d)~profile of the transmitted beam calculated taking into account the imperfections of the fabricated structure.}
\end{figure}

\begin{figure}[tbh]
	\centering
	\includegraphics{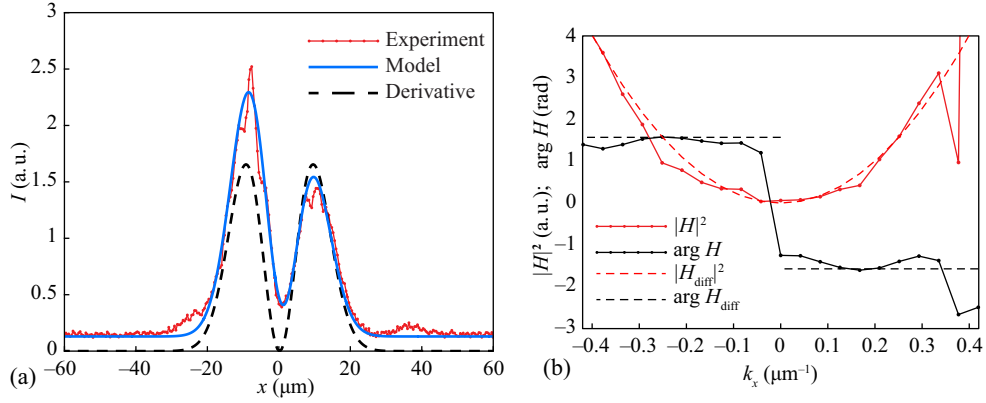}
	\caption{\label{fig:5}(a) Cross sections of the transmitted beam profile: exact derivative (dashed black curve), model (solid blue curve), experiment (red curve with dot markers); (b) intensity (squared absolute value) and phase (argument) of the transfer functions of an exact differentiator (dashed curves) and of the estimated transfer function of the fabricated structure (solid curves with dot markers).}
\end{figure}

The results of the experimental investigation of the fabricated structure is shown in Figs.~\ref{fig:4} and~\ref{fig:5}.
As mentioned above, since the parameters of the fabricated grating slightly differ from the design values, these differences have to be compensated by tuning the frequency and/or the incidence angle $\theta$ of the impinging beam (see Fig.~\ref{fig:3}). 
In the experiment, the measurements were performed at the incidence angle $\theta = 1.04^\circ$ and the wavelength $\lambda = 625\nm$, which correspond to the minimum of the transmittance necessary for performing the differentiation operation. 
The profiles of the incident Gaussian beam and the transmitted beam are shown in Figs.~\ref{fig:4}(a) and~\ref{fig:4}(b).
The shape of the incident beam is close to Gaussian $P_{\rm inc} (x,y)=\exp \left\{-\left(x^2 + y^2 \right) / \left(2\sigma^2 \right)\right\}$ at $\sigma = 9.52\um$.
Figure~\ref{fig:4}(c) shows the profile of the analytically calculated derivative of the Gaussian beam $\partial P_{\rm inc} (x,y) / \partial x  \sim x \exp \left\{-\left(x^2 + y^2 \right)/\left(2\sigma^2 \right)\right\}$. 
The shape of the transmitted beam is in good agreement with the exact derivative [Figs.~\ref{fig:4}(b) and~\ref{fig:4}(c)].

The discrepancies between the shape of the transmitted beam and the analytically calculated derivative are caused by imperfections of the fabricated structure and the CCD noise.
Let us consider a model that takes into account these effects and enables calculating the profile of the transmitted beam corresponding to the imperfect structure.
The transfer function of a perfect differentiator [Eq.~\eqref{eq:5}] is linear with respect to $k_x$. 
At the central angle of incidence $\theta = \arcsin \left( k_{x'0} / k_0 \right)$, the transmission coefficient of the differentiator [Eq.~\eqref{eq:4}] vanishes.
If the geometry of the structure is not strictly periodic (height and/or the period of the grating vary), or if the structure is not strictly lossless, the transmission coefficient becomes non-zero at the considered angle of incidence. 
In the first approximation, it leads to the addition of a constant term to the transfer function:
\begin{equation}
\label{eq:6}
H\left(k_x \right) \sim a + {\rm i}k_x,
\end{equation} 
where the complex number $a$ can be fitted from the experimental data. 
Figure~\ref{fig:4}(d) shows the profile of the beam calculated using the transfer function of Eq.~\eqref{eq:6} at $a=0.012+0.026\ii\um^{-1}$. 
Similarly to Eq.~\eqref{eq:5}, in Eq.~\eqref{eq:6} the tilde notation assumes that a multiplicative constant is omitted.
For the fabricated structure, the estimated value of this constant amounts to $\alpha' = 8.8\um$.
Figure~\ref{fig:5}(a) shows the cross sections of the transmitted beam obtained in the experiment (red curve with dot markers), of the analytically calculated derivative (dashed black curve), and of the beam profile calculated using the model of Eq.~\eqref{eq:6} (solid blue curve). 
From Figs.~\ref{fig:4}(b),~\ref{fig:4}(c) and~\ref{fig:5}(a), it is evident that the model of Eq.~\eqref{eq:6} allows us to describe the shape of the transmitted beam with good accuracy and to explain its deviation from the exact derivative.

On the basis of the experimental data, we also estimated the transfer function of the fabricated diffraction grating using the following approach. 
Using the estimate of Eq.~\eqref{eq:6}, the profile of the transmitted beam $P_{\rm tr} \left(x ,0\right)$ was calculated [see Eq.~\eqref{eq:A8} in the Appendix].
The amplitude of the calculated profile was replaced by the measured one $P_{\rm tr, meas}$, while the phase was retained: $\hat{P} = P_{\rm tr, meas} \left(x, 0\right) \cdot \exp \left\{\ii \arg P_{\rm tr} \left(x ,0\right)\right\}$. 
Then, using the discrete Fourier transform, complex spectrum of the transmitted beam ${\mathcal F}\left\{\hat{P}\right\}$ was calculated. Finally, this spectrum was divided by the known spectrum of the incident Gaussian beam $G\left(k_x \right)\sim \exp \left\{-k_x^2 \sigma^2/2\right\}$, yielding the complex transmission coefficient of the structure and, consequently, the sought-for estimate of the transfer function of the fabricated grating. 
The obtained estimate was further refined using the Gerchberg--Saxton iterative algorithm, in which the measured amplitudes of the transfer function and of the transmitted beam were used as constraints. 
The obtained transfer function is close to the transfer function of a perfect differentiator of Eq.~\eqref{eq:5} [see Fig.~\ref{fig:5}(b)].

Let us note that in contrast to metasurface-based structures considered in~\cite{Silva:2014:science, Pors:2015:nl}, the fabricated structure is more compact and does not require the utilization of the optical setup of the Fourier correlator containing one or two lenses. In comparison with these works, the fabricated grating of the present work provides significantly better differentiation quality. In particular, the transfer function of the fabricated structure is closer to the transfer function of an exact differentiator: compare Fig.~\ref{fig:5}(b) of the present work and Fig.~4(c) of~\cite{Pors:2015:nl}.

\section{Conclusion}
We investigated the transformation of the transverse profile of an optical beam by a guided-mode resonant grating both analytically and experimentally.
We showed that this transformation can be described in terms of linear systems theory by the transfer function associated with the resonant grating. 
By analyzing the general form of the transfer function, we deduced the conditions enabling spatial differentiation of the transverse profile of the incident beam. 
The first-order spatial differentiation was demonstrated experimentally for a micrometer-scale incident Gaussian beam.
On the basis of the measured profiles of the incident and transmitted beams, we estimated the transfer function of the fabricated grating, which was shown to be in good agreement with the transfer function of an exact differentiator.

\appendix
\section*{\label{appendix}Appendix: Transfer function of a diffraction grating}

Here, we describe the diffraction of an obliquely incident optical beam on a diffraction grating, assuming that the vector describing the propagation direction of the beam lies in the \textit{x'z'} plane (see Fig.~\ref{fig:1}). 
The coordinate system $(x, y, z)$, which is associated with the incident beam, is related to the coordinate system $(x', y, z')$, which is associated with the grating, by the following expressions:
\begin{equation}
\label{eq:A2}
 \begin{aligned}
x &= x'\cos \theta + z'\sin \theta, \\ 
z &=-x'\sin \theta + z'\cos \theta.
\end{aligned}
\end{equation}
Here, we assumed that the origins of these coordinate systems coincide.
According to Eq.~\eqref{eq:A2}, the wave vector with components $\lt(k_{x'}, k_y, k_{z'}\rt)$ in the coordinate system associated with the grating has the following form in the coordinate system associated with the incident beam: 
\begin{equation}
\label{eq:A0}
\lt(k_x, k_y, k_z\rt)=\lt(k_{x'} \cos \theta + k_{z'} \sin \theta, k_y, k_{z'} \cos \theta - k_{x'} \sin \theta\rt).
\end{equation}

Let us represent the beam as a plane-wave expansion. 
%The plane waves in this expansion possess non-zero tangential components of the wave vector ($k_x$ and $k_y$). 
In the present work, we use the following basis of TM- and TE-polarized plane waves defined in the grating coordinate system $(x', y, z')$:
\begin{equation}
\label{eq:A1}
\pmb\Psi_{\rm TM} =
\frac{1}{k_0 k_{z'} } 
\begin{bmatrix}
 k_0 k_{z'} \\ 0 \\ -k_0 k_{x'} \\ -k_{x'} k_y \\ k_{x'}^2 +k_{z'}^2 \\ -k_y k_{z'} 
\end{bmatrix}
\ee^{\ii k_{x'} x'+\ii k_y y+ \ii k_{z'} z'},\, \, \, \, \, \, \, \, 
{\pmb\Psi}_{\rm TE} =\frac{1}{k_0 k_{z'} } 
\begin{bmatrix}
 k_{x'} k_y  \\ -\lt(k_{x'}^2 +k_{z'}^2 \rt) \\ k_y k_{z'}  \\ k_0 k_{z'} \\ {0} \\ -k_0 k_{x'} 
\end{bmatrix}
\ee^{\ii k_{x'} x'+\ii k_y y+ \ii k_{z'} z'},
\end{equation}
where $k_{x'}^2 + k_y^2 + k_{z'}^2 = k_0^2$ and the column vectors ${\pmb\Psi}$ contain the components of the electromagnetic field: ${\pmb\Psi} =\lt[E_{x'} \; E_y \; E_{z'} \; H_{x'} \; H_y \; H_{z'} \rt]^{\rm T}$.
Let us note that at $k_y =0$ Eq.~\eqref{eq:A1} describes the ``conventional'' TM- and TE-polarized plane waves with $\tilde{{\pmb\Psi}}_{\rm TM} =\lt[E_{x'} \; 0  \; E_{z'}  \; 0  \; H_y  \; 0\rt]^{\rm T} $, 
$\tilde{{\pmb\Psi}}_{\rm TE} = \lt[0  \; E_y  \; 0  \; H_{x'}  \; 0  \; H_{z'} \rt]^{\rm T}$.
Let us also note that the basis in Eq.~\eqref{eq:A1} is orthogonal, and a plane wave of arbitrary polarization can be represented as a superposition of the waves of Eq.~\eqref{eq:A1}.

We will consider the case when the incident beam is polarized along the $x$~axis. 
In this case, the electric field has zero $E_y$ component, and the $E_x$ component in the plane $z = 0$ is described by a certain function $P_{\rm inc} (x, y)$. Let us define the beam spectrum as the Fourier transform of the $E_x$ field component at $z = 0$:
\begin{equation}
\label{eq:A3}
 G\lt(k_x, k_y\rt)=\frac{1}{(2\pi)^2 } \iint P_{\rm inc} \lt(x, y\rt)\ee^{-\ii k_x x - \ii k_y y} \dd x \dd y.
\end{equation}

According to Eqs.~\eqref{eq:A1} and~\eqref{eq:A2}, the complex amplitude of the $E_x$ component of the wave ${\pmb\Psi}_{\rm TM}$ equals $\cos \theta - k_{x'} \sin \theta / k_{z'}  = k_z / k_{z'}$. Therefore, the wave $(k_{z'} / k_z ) {\pmb\Psi} _{\rm TM}$ has unity-amplitude $E_x$ field component. Such plane waves can be used as a basis for the expansion of the incident beam field:
\begin{equation}
\label{eq:A4}
{\pmb\Psi}_{\rm inc} \lt(x,y,z\rt) = \iint G\lt(k_x, k_y\rt) \frac{k_{z'}}{k_z} {\pmb\Psi}_{\rm TM} \dd k_x \dd k_y.
\end{equation}
where ${\pmb\Psi}_{\rm TM}$ contains $k_{x'}$ and $k_{z'}$ which are related with $k_x$ and $k_z$ through Eq.~\eqref{eq:A0}.
It is easy to verify by direct substitution that the expansion of Eq.~\eqref{eq:A4} provides fulfillment of Eq.~\eqref{eq:A3} at $E_x\lt(x,y,0\rt)=P_{\rm inc}\lt(x, y\rt)$. Let us note that the choice of the expansion basis of Eq.~\eqref{eq:A1} allows us to expand the incident beam in TM-polarized waves only. However, in the case when the incident beam is polarized along the $y$ axis, the incident beam is represented as a linear combination of both TE- and TM-polarized waves.

Equation~\eqref{eq:A4} reduces the problem of diffraction of a beam ${\pmb\Psi}_{\rm inc}$ to multiple problems of diffraction of plane waves ${\pmb\Psi}_{\rm TM}$ on a diffraction grating. Note that the solution of these problems (the reflected and transmitted fields) contains waves of both polarizations. In particular, the transmitted field in the case of incidence of a TM-polarized plane wave ${\pmb\Psi}_{\rm TM}$ has the form
\begin{equation}
\label{eq:A5}
T_{\rm same} \lt(k_{x'},k_y \rt){\pmb\Psi}_{\rm TM} + T_{\rm cross} \lt(k_{x'},k_y \rt){\pmb\Psi}_{\rm TE},
\end{equation}
where $T_{\rm same} \lt(k_{x'},k_y \rt)$ is the complex amplitude of the TM-polarized zeroth transmitted diffraction order in the case of incidence of a TM-polarized wave and $T_{\rm cross} \lt(k_{x'},k_y\rt)$ is the complex amplitude of the cross-polarized zeroth diffraction order. According to Eq.~\eqref{eq:A5}, the resulting transmitted field in the case of beam incidence has the form
\begin{equation}
\label{eq:A6}
{\pmb\Psi}_{\rm tr} (x,y,z) 
= \iint G\lt(k_x,k_y \rt)\frac{k_{z'}}{k_z} \lt[T_{\rm same} \lt(k_{x'},k_y \rt){\pmb\Psi} _{\rm TM} +T_{\rm cross} (k_{x'},k_y ){\pmb\Psi}_{\rm TE} \rt] \dd k_x \dd k_y  .
\end{equation}
Approximate expressions for the transmission coefficients $T_{\rm same} \lt(k_{x'}, k_y \rt)$ and $T_{\rm cross} \lt(k_{x'}, k_y \rt)$ were given in~\cite{my:Bykov:2017:oe}. According to these expressions, the coefficient $T_{\rm cross} \lt(k_{x'},k_y \rt)$ can be neglected at small $k_y$ as a quantity of a higher order of smallness than the coefficient $T_{\rm same} \lt(k_{x'}, k_y\rt)$. In this case, the transmitted field takes the following form:
\begin{equation}
\label{eq:A7}
{\pmb\Psi}_{\rm tr} (x,y,z) = \iint G\lt(k_x,k_y \rt)\frac{k_{z'} }{k_z } T_{\rm same} \lt(k_{x'}, k_y \rt) {\pmb\Psi}_{\rm TM} \dd k_x \dd k_y.
\end{equation}
The $E_x$ component of the transmitted beam in the plane $z = 0$ can be written as
\begin{equation}
\label{eq:A8}
 P_{\rm tr} (x, y)=\iint G\lt(k_x,k_y \rt)T_{\rm same} \lt(k_{x'},k_y \rt)\ee^{\ii k_x x +\ii k_y y} \dd k_x \dd k_y.
\end{equation}
It follows from Eq.~\eqref{eq:A8} that the transformation of the profile of the incident beam upon transmission through the diffractive structure is described by multiplication by the term $T_{\rm same} \lt(k_{x'},k_y \rt)$ in the Fourier space. 
This result shows that the diffraction grating is described by a linear system with the transfer function of Eq.~\eqref{eq:1}. Note that in Eq.~\eqref{eq:1}, the $k_{x'}$ wave vector component is expressed through the components $k_x$ and $k_z$ using the coordinate transformation of Eq.~\eqref{eq:A0} and free-space dispersion equation $k_x^2 + k_y^2 + k_z^2 = k_0^2$, where $k_z < 0$.

Let us note that with a number of additional assumptions, a transfer function similar to that of Eq.~\eqref{eq:1} can also be obtained for the case of a beam polarized along the $y$ axis.

\section*{Funding}
Russian Science Foundation (14-19-00796) (experiment),
Russian Foundation for Basic Research (16-29-11683) (design of the grating),
and Federal Agency of Scientific Organizations (agreement \textnumero 007-GZ/Ch3363/26) (analysis of the grating transfer function).

\end{document}